\documentclass[twocolumn,preprintnumbers,amsmath,amssymb]{revtex4}
\usepackage{graphicx}% Include figure files
\usepackage{dcolumn}% Align table columns on decimal point
\usepackage{amsmath}
\usepackage{bm}% bold math

\begin{document}
\title{Graphene on hexagonal lattice substrate: Stress and Pseudo-magnetic field}
\author{M. Neek-Amal}
\author{F.M. Peeters}
\affiliation{Department of Physics, University of  Antwerpen, Groenenborgerlaan 171, B-2020 Antwerpen, Belgium}
%\author{A. K. Geim}
%\affiliation{School of Physics and Astronomy, University of Manchester, Manchester, M13 9PL, UK}
\date{\today}

\begin{abstract}
 Moir\'e patterns in the pseudo-magnetic field and in the strain profile
of graphene (GE) when put  on top of a hexagonal lattice substrate
are
predicted from elasticity theory. %which are confirmed by atomistic simulations.
 The van der Waals (vdW) interaction between GE and the
substrate induces out-of-plane
 deformations in graphene  which results in a strain field, and consequently in a
  pseudo-magnetic field. When the misorientation angle is
about 0.5$^o$ a three-fold symmetric strain field is realized that
results in a pseudo-magnetic field very similar to the one proposed
by F. Guinea,  M. I. Katsnelson, and A. K. Geim [Nat. Phys. {\bf6},
30 (2010)]. Our results show that the periodicity and length of the
pseudo-magnetic field can be tuned in GE by changing the
misorientation angle and substrate adhesion parameters and a
considerable energy gap (23\,meV) can be obtained due to
out-of-plane deformation of graphene which is in the range of recent
experimental measurements~(20-30\,meV).

\end{abstract}
\maketitle Stacking different two dimensional materials with
slightly different lattice structures on top of each other results
in a new superlattice structure which is called Moir\'e pattern. The
van der Waals (vdW) interaction between different 2D-crystals such
as graphene (GE), hexagonal boron nitride (h-BN), and molybdenum
disulfide (MOS$_2)$ results in a multilayer
heterostructure~\cite{vdw}. The resulting hexagonal Moir\'e pattern
in graphene on top of other hexagonal lattice substrates affects the
electromechanical properties of graphene. For example, hexagonal
boron nitride  has turned out to be an ideal dielectric substrate
which is atomically flat and improves graphene's mobility by more
than two orders of magnitude~\cite{Dean,Wei}. The B-N bond length is
close to that   of C-C with only a very small (1.6$\%$-2$\%$)
lattice mismatch~\cite{4,javad} which results in the appearance of a
Moir\'e pattern (MP) when GE is put on top of BN. It was found that
GE flakes can align with the underlying h-BN lattice within an error
of less than 0.05$^o$~\cite{4,6}. Ab-initio and semi-empirical   van
der Waals  studies showed that the interaction between  GE flakes
and the h-BN substrate is similar to that of a GE-GE stacked
structure~\cite{SACH}. On the other hand the different
electronegativity of B, N and C atoms leads to a non-uniform
attractive force distribution over GE.

Non-uniform strain in GE results in a pseudo-magnetic field and
consequently results in the opening of an energy
gap~\cite{NatPhys,neek2013}. Earlier density functional theory
calculations assumed lattice matching between GE and h-BN which
induces in-plane strain and opens a gap in GE's spectrum of
50-60~meV~\cite{SACH}. But recent experiments found a gap in the
range of 20-30\,meV~\cite{geimarXiv,science2013}.  In this letter,
we first develop a general theory for GE over a hexagonal lattice
substrate and show that the induced strain has triangular symmetry
resulting in interesting pseudo-magnetic field patterns which vary
with the misorientation angle. Then, as an example, we concentrate
on the h-BN-lattice induced deformation of the GE lattice using
atomistic simulations and compare with our analytic results. Using
experimental height deformation~\cite{pekhers} as an input in our
analytic theory we found that the pseudo-magnetic field modulation
amplitude can be of order 1 Tesla for misorientation less than
1$^{o}$. The latter results in an energy gap of  about 23\,meV.
\begin{figure}
\begin{center}
\includegraphics[width=1.0\linewidth]{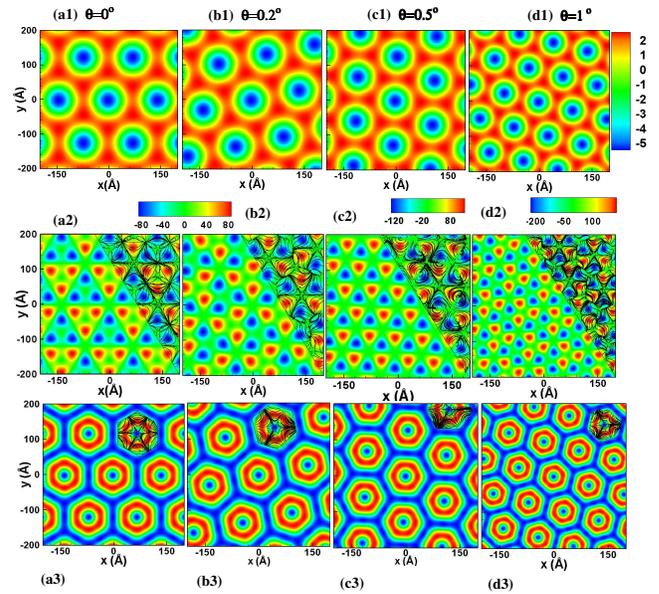}
 \caption{ (a1-c1) Height deformations ($\Delta h/ h_0$)  of GE over h-BN sheet, i.e. Eq.~(1),
  for different misorientation angle with lattice mismatch 1.7$\%$. (a2-c2) The corresponding induced magnetic field (per $h_0^2$), Eq.~(\ref{Magnetic2}).
 Typical streamlines of gauge field vector are shown in the
corner of each panel(a3-c3). The strain tensor eigenvalue, i.e.
$\epsilon_+$ (which is equivalent to the absolute value of the gauge
field).  %(a4-c4) The opened gap for each panel is calculated
%according to the Landau level quantization, i.e. $\propto \sqrt{B}$.
\label{fig1}}
\end{center}
\end{figure}
\emph{{The model.}}
 The mismatch between the honeycomb lattice structures of GE and a hexagonal lattice (e.g. h-BN) leads to long wavelength  Moir\'e
 patterns. For a given lattice mismatch and misorientation between
 GE and the substrate lattice we expect  that the GE sheet is attracted to the h-BN
 substrate (the adhesion energy for GE/h-BN is about 30-50\,meV per atom~\cite{SACH,APL2014}). For a lattice mismatch $\delta$ and misorientation angle $\theta$
 (with the zig-zag (zz) direction along the x-axis) the deformation of the lattice   due
 to vdW adhesion is $\Delta h(x,y)$ which also depends on   the vdW adhesion strength $\varepsilon$.
 We found that the symmetry of the  out-of-plane deformation
 in the GE lattice is similar to the MP structure~\cite{Proof}.
  Therefore the Fourier transform (FT) of $\Delta h$ should correspond to six Moir\'e pattern
vectors~\cite{4,APL2014,Wallbank}, i.e. $\vec{G}_{m}=
\Re_{\phi_m}\vec{G}_0$ with $m=0,1,..5$ where
$\vec{G}_0=(\hat{1}-\frac{1}{1+\delta)}\Re_{\theta})(0,2\kappa)$
with  $\kappa=\frac{2\pi}{3a_{0}}$ and $\Re_{\phi_m}$ (and
$\Re_{\theta}$) is the rotation matrix about the $z$-axis over an
angle $\phi_m=\frac{2 \pi m}{6}$ (and $\theta$) where $a_0=a_{CC}$
is the C-C bond length.~The height deformation of GE can be
generally written as
%\begin{equation}
    $\Delta h=h_0 \sum_m e^{i \vec{G}_m.\vec{r}}$, %\label{h1}
%   \end{equation}
where $h_0$ is the amplitude of the deformation. For $\theta<<1^{o}$
we simplify the modulation function  as
\begin{eqnarray}
\frac{\Delta h}{2 h_0}=\cos [\vec{r}. \vec{G_0} ]+2\cos
[\frac{\vec{r}. \vec{G_0}}{2}] \cos [\frac{\sqrt{3}}{2}\vec{r}\times
\vec{G_0}].\label{h2}
   \end{eqnarray}
The elements of the strain tensor  can be found using
 $\epsilon_{\alpha\beta}= \frac{1}{2}\partial_{\alpha}h
\partial_{\beta}h$. The x-component of the strain tensor is a periodic
 function and shows mirror symmetry along the zz direction while the y-component shows three fold symmetry with large
peaks on the hexagonal sites. The shear component shows two fold
symmetry and is three times smaller. Diagonalising the strain tensor
gives the principal axis with eigenvalues
\begin{equation}\epsilon_{\pm}=\frac{1}{2}[\epsilon_{ii}\pm|\vec{A}|]
\end{equation}
 where $\vec{A}$ is the gauge vector corresponding to the lattice
deformation~\cite{geim}. Surprisingly we found that  $\epsilon_-=0$
(since $\epsilon_{xx}\epsilon_{yy}=\epsilon_{xy}$) and
$\epsilon_+=|\vec{A}|$ having MP properties. For a  two dimensional
material  we found  that the corresponding eigenvectors
have an  angle
\begin{equation}
\Phi_{-}=\pi-tan^{-1}(\frac{\epsilon_{xy}}{\epsilon_{xx}}),\Phi_+=\pi-tan^{-1}(\frac{\epsilon_{xy}}{\epsilon_{yy}})\end{equation}
with respect to the zz-direction. Since the eigenvalue
$\epsilon_-=0$, we conclude that the stress along the corresponding
eigenvector results in no lattice deformation.

The low energy electronics of the deformed GE can be obtained from
the Dirac equation after inserting the modified hopping parameters
from the  tight-binding model which are now a function of the atomic
positions $t(\textbf{r})$~\cite{neek2013}. Rewriting the Dirac
Hamiltonian in the effective mass approximation introduces now the
strain induced effective gauge field
$\vec{A}=\frac{2\beta_0\hbar}{3a_0
e}(\epsilon_{xx}-\epsilon_{yy},-2\epsilon_{xy})$ where $\beta_0$
($\sim$2-3) is a constant (the strain due to the out of plane
displacements~\cite{geim}). Using  two components of the effective
gauge field in the unit $\tau=8\kappa^2 \beta_0 \hbar {h_0}^2/3a_0e$
are
%\begin{widetext}
%\begin{eqnarray*}
%%$\begin{align}
%        A_x&=&2 [\theta  \cos (\sqrt{3}\kappa \chi_1) \sin
%(\kappa\chi_2)+\theta \sin (2\kappa\chi_2)-\sqrt{3} \delta \cos
%(\kappa\chi_2 )   \sin (\sqrt{3}\kappa \chi_1)]^2\\\nonumber &&-2
%[\delta \cos (\sqrt{3} \kappa\chi_1) \sin (\kappa\chi_2 )+\delta
%\sin (2\kappa\chi_2)+\sqrt{3} \theta  \cos (\kappa\chi_2) \sin
%(\sqrt{3}\kappa \chi_1)]^2\\
%        A_y&=&4 [\theta  \cos (\sqrt{3}\kappa \chi_1) \sin (\kappa\chi_2
%)+\theta \sin (2\kappa\chi_2)-\sqrt{3} \delta  \cos (\kappa\chi_2
%)\\\nonumber &&
%   \sin (\sqrt{3}\kappa \chi_1))][\delta  \cos (\sqrt{3}\kappa \chi_1) \sin (\kappa\chi_2)+\delta  \sin (2\kappa\chi_2
%)+\sqrt{3} \theta  \cos (\kappa\chi_2) \sin (\sqrt{3}
%\kappa\chi_1)],
%%\end{align}
%%\end{subequations}
%\end{eqnarray*}
%   \end{widetext}
%where $2\kappa\chi_1=\vec{r}\times \vec{G_0}$ and
%$2\kappa\chi_2=\vec{r}. \vec{G_0}$.
we can find the corresponding curvature induced magnetic field
perpendicular to the $x-y$ planes in units of $2\kappa \tau$ given
by
\begin{widetext}
\begin{eqnarray}
 B&=&4\sqrt{3}\theta\omega_1 \cos
(\sqrt{3}\kappa \chi_1) \sin (\sqrt{3}\kappa \chi_1) \cos
^2(\kappa\chi_2 )-4\delta\omega_2  [\sin (\kappa\chi_2 ) \cos
^2(\sqrt{3}\kappa \chi_1) +\sin (2\kappa\chi_2) \cos (\sqrt{3}\kappa
\chi_1 )+\\\nonumber &&-3 \sin (\kappa\chi_2 ) \sin
^2(\sqrt{3}\kappa \chi_1 )] \cos (\kappa\chi_2 )+
4\sqrt(3)\theta\omega_1  \sin (\kappa\chi_2 ) [\cos (\sqrt{3}\kappa
\chi_1) \sin (\kappa\chi_2)+\sin (2\kappa\chi_2 )] \sin
(\sqrt{3}\kappa\chi_1 )+\\\nonumber &&\cos (2\kappa\chi_2 ) [\delta
\omega_2 \cos (\sqrt{3}\kappa \chi_1) \sin (\kappa\chi_2)+\delta
\omega_2 \sin (2\kappa\chi_2 )- 4\sqrt{3}\theta \omega_1 \cos
(\kappa\chi_2 ) \sin (\sqrt{3}\kappa\chi_1)],\label{Magnetic2}
      \end{eqnarray}
   \end{widetext}
where $2\kappa\chi_1=\vec{r}\times \vec{G_0}$,
$2\kappa\chi_2=\vec{r}. \vec{G_0}$ $\omega_1=3\delta^2-\theta^2$ and
$\omega_2=3\theta^2-\delta^2$. The corresponding stress tensor is
given by $\sigma_{ij}=\lambda \delta_{ij}\epsilon_{ii}+2\mu
\epsilon_{ij}$, where $\lambda$ and $\mu$ are the Lam\'{e}
parameters that determine the stiffness of the material. It is
interesting to note that the gauge field is proportional to the main
element of the stress tensor, i.e.
$\sigma_+=\sigma_{ii}=2K\epsilon_+$ with $K=12.3\,eV\AA^{-2}$~being
the 2D bulk modulus of GE~\cite{JAN}. However, to find the energy
levels and localized states  in the K valley one can solve the
following equation for the eigen-energy E:
\begin{equation}
v_F^2[{\vec{\Pi}}^2\pm\,i\,e\,(\vec{p}\times
\vec{A})]\psi_{K_\pm}=E^2\psi_{K_\pm}\label{Dirac}
\end{equation}
 where $v_F$ is
the Fermi velocity, $\pm$ refer to the A and B-sites in the GE
lattice and $\vec{\Pi}=\vec{p}+e\vec{A}$. %\textbf{By relaxing the GE lattice over BN sheet using atomistic simulations~\cite{APL2014} we found that the maximum of $\delta h$ is much smaller than 1\AA.}

\begin{figure}
\begin{center}
\includegraphics[width=1.0\linewidth]{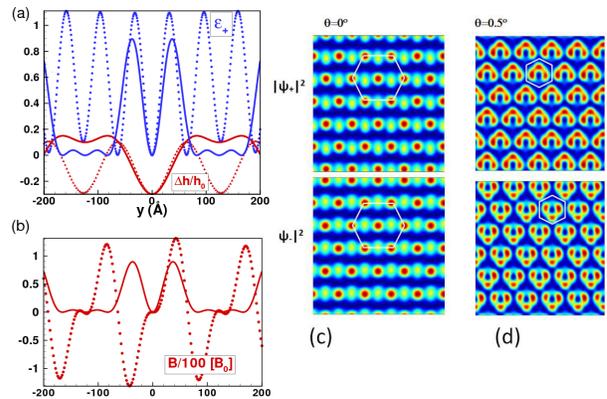}
 \caption{  (a) The red lines are two cross sections along the ac direction, i.e. $x$=0, from a2 (solid-line  $\theta=0^{o}$)
 and c2 (dashed-line  $\theta=0.5^{o}$) panels (for pseudo-magnetic field)  in
 Fig.~\ref{fig1}. %The corresponding energy gap  are shown by blue
% color.
(b) The red  lines are two cross sections along ac direction, i.e.
$x$=0, from a1 (solid-line $\theta=0^{o}$) and c1
 (dashed-line $\theta=0.5^{o}$) panels in
 Fig.~\ref{fig1} for height deformation. The corresponding strain
 eigenvalues (and absolute value of gauge vector) are shown by blue
 color, i.e. a3 (solid-line $\theta=0^{o}$) and c3 (dashed-line $\theta=0.5^{o}$) panels.
(c,d) The solution of continuum Dirac equation (Eq. (\ref{Dirac}))
for
 GE over a substrate with hexagonal lattice structure, $\psi_{\pm}$ are the  wave functions over A- and B-sites for low energy
 E=0.3\,eV.
\label{fig2}}
\end{center}
\end{figure}

In Fig.~\ref{fig1} we collect all the results for the height
deformations (labeled by 1),~ pseudo-magnetic field (labeled by 2 in
units of $B_0={h_0}^{2}\, T\AA^{-2}$), the $\epsilon_+$ in arbitrary
units (equivalently, the absolute value of gauge field labeled by 3)
 for different misorientation
angles, $\theta=0^{o},0.1^{o},0.5^{o}$ and $1^{o}$ from left to
right, respectively. The MP structure appears in $\Delta h$ and
$\epsilon_{+}$ panels with smaller period for larger $\theta$.
Notice that the main MP hexagon in the central part of a1 (a3) for
$\Delta h$ ($\epsilon_+$) is rotated by an angle
$\varphi=tan^{-1}\frac{\sin(\theta)}{\cos(\theta)+\delta-1}$
clockwise and is scaled by
$L=a_0[2(1-\delta)(1-cos(\theta))+\delta^2]^{-1/2}$. It is seen that
for $\theta=0.5^{o}$ (c1) the height profile is rotated over $\pi/2$
with respect to the zero angle (a1) result. Notice that the symmetry
of the pseudo-magnetic field profiles vary with $\theta$. The
$\theta$ angle which  reproduces the pseudo-magnetic field results
of Ref.~\cite{NatPhys} is $\theta=0.5^{o}$ (i.e. triaxially stressed
GE). The latter is related to the angle
$\varphi(\theta=0.5^{o})\simeq\pi/6$.
 Notice that the magnetic field profile for $\theta=0^{o}$
 is completely different from  what one expects intuitively, i.e. there is
no hexagonal symmetry as in the $0.5^{o}$ case. Here, the obtained
pseudo-magnetic field is only due to out-of-plane deformation. If
in-plane triaxial stress is applied it will increase the
pseudo-magnetic field further~\cite{NatPhys,neek2013}. The change in
the pseudo-magnetic field with  $\theta$ is  a    very promising
method to tune the electronic gap with the lattice misorientation.
 The profiles in $\epsilon_+$ do not have a  sinusoidal shape along particular
directions, e.g. in c3 along ($1,\sqrt{3}$). The blue hexagonal
patterns in a3-d3 are the minimum value of the strained part in GE which
are connected to each other with the MP structure. However the highest
strained regions (red hexagons)  show smaller hexagonal
patterns. Between these two regions (blue-bigger hexagon and
red-smaller hexagons) there are moderate green hexagons. The latter
effect is closely related to the recent PeakForce tapping atomic force
microscopy  measurement of the mechanical properties of GE/h-BN
where the tip is sensitive to the  strain
distribution in graphene~\cite{geimarXiv}.
% The estimated energy gap also varies strongly with
%$\theta$ which should affect the conductivity in GE.
The corresponding cross sections of panel a2 are shown by blue-solid
lines for the pseudo-magnetic field in Fig.~\ref{fig2}(a). In
Fig.~\ref{fig2}(b) we show  cross sections of $\Delta h$ and
$\epsilon_+$ along the ac direction for $\theta=0$ (a1, a3 blue
solid lines) and $0.5^{o}$ (c1, c3 red dashed lines).

In Figs.~\ref{fig2}(c,d) we depict the solution of Eq.
(\ref{Dirac}), i.e. the probability density of wave function, at the
 A- and B-lattice ($\psi_{\pm}$) for low energy E=0.3\,eV where
in (c) $\theta=0^{o}$ and in (d) $\theta=0.5^{o}$. The hexagons
indicate the corresponding MP. Notice that the probability densities
are completely different even for small misorientation angle.

Our analytical MP scale free results presented in
Figs.~\ref{fig1}(a1-c1) are in good agreement with recent
experiments on epitaxial grown graphene on h-BN~\cite{NAtMat2013}.
From the experimental measured amplitude $\Delta h\simeq$ 40\,pm and
using $h_0\sim\Delta h/5.0$, the pseudo-magnetic field in
Fig.~\ref{fig1}(a2) varies in the [-0.5,0.5]\,T range (i.e.
$B_0$=6.4\,mT). We estimated the energy gap using the classical
approximation for  relativistic electrons in the presence of
magnetic field where the Landau level energy sequence can be written
as

\begin{equation}
E_N=\pm\sqrt{2\,e\,\hbar\,v_F^2\, B\, N}\approx \pm 400 k_B
\sqrt{B\,N}\label{EN}
\end{equation}
The energy gap can be approximated as $\Delta\simeq 33.3 \,meV$ for
B=1\,T. The experimentally observed strain distribution in graphene
over h-BN was found to be different for commensurate and
incommensurate states~[10] which can be equivalent to the transition
from a non-uniform to an uniform strain distribution in graphene.
The non-uniform strain in graphene
 results in an opening of a gap in some graphene over h-BN~[9], e.g. atomic force microscopy
 measurement predicts a gap of $\sim$30\,meV~[10] and 20\,meV ~[11]. The experimentally
 measured amplitude $\Delta h\simeq$ 40\,pm results in B$\sim$0.5\,T and by using Eq.~(6) in our model we find
$\Delta\approx$\,23 meV which is comparable with available
experimental results.

\begin{figure}
\begin{center}
\includegraphics[width=0.7\linewidth]{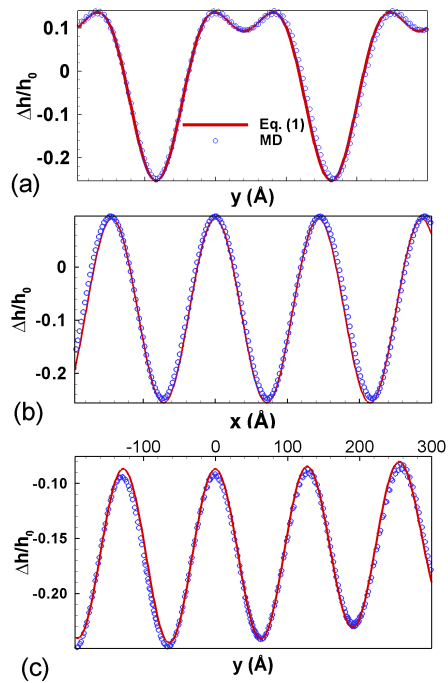}
\caption{(Color online)  The height deformation from molecular
dynamics relaxation for GE/h-BN sheet and corresponding analytical
results  from Eq.~(1). The lattice mismatch is $\delta=1.7\%$ and
two cross sections in (a,b) correspond to $\theta=0^{o}$ and in (c)
$\theta=0.5^{o}$(circular symbols were taken from
Figs.~\ref{fig1}(a1,c1)). \label{figMD}}
\end{center}
\end{figure}

Finally, in order to determine $h_0$ and find the scale $B_0$ in the
previous analysis and have an independent check for the above theory
we use a Lennard-Jones potential containing both the short range
repulsive and long range attractive nature of the interaction
between two particles i.e.
$u(r)=4\varepsilon[(\sigma/r)^{12}-(\sigma/r)^6 )]$, where r is the
distance between two atoms, $\varepsilon$ is the depth of the
potential well, and $\sigma$ is the distance at which the potential
becomes zero. To model the interaction between B, N and C atoms, we
adjust the LJ parameters using the equations
$\varepsilon=\sqrt{\varepsilon_i \varepsilon_j}$ and
$\sigma=(\sigma_i+\sigma_j)/2$ where $i,j$  refer to B, N or C and
where $\sigma_C$ =3.369\,\AA~, $\sigma_B$ =3.453\,\AA,~$\sigma_N$
=3.365\,\AA~ and $\varepsilon_C$=2.635 meV, $\varepsilon_B$=4.16
meV, and $\varepsilon_N$=6.281 meV~[13]. We relaxed the
computational unit cell of GE/h-BN (for $\theta=0^{o}$ and
$\theta=0.5^{o}$) using molecular dynamics relaxation with the
conjugated gradient algorithm (with periodic boundary condition).
The GE sample and h-BN layer consist of 250632 C and 242208 B and N
atoms, respectively (i.e. $N_{tot}=$ 492840). %The potential well in
%the used LJ potential is set to $\varepsilon_C$=2.635\,meV,
%$\varepsilon_B$=4.16\,meV, and $\varepsilon_N$=6.281\,meV~\cite{8,9}
%and $\sigma_C$ =3.369\,\AA,~ $\sigma_B$ =3.453\,\AA,~$\sigma_N$
%=3.365\,\AA.~To model the interaction between B, N and C atoms, we
%adjust the LJ parameters using the equations
%$\varepsilon_0=\sqrt{\varepsilon_i \varepsilon_j}$ and
%$\sigma=(\sigma_i+\sigma_j)/2$ where $i,j$ refer to B, N or C. For
%the interaction between the C atoms of GE we employed the AIREBO
%potential which is suitable for simulating hydrocarbons as
%implemented in the LAMMPS package~\cite{AIREBO}. More details of the
%used method can be found in our previous work~\cite{APL2014}.
 After
relaxation we found that the average C-C bond length in GE is
$a_{CC}=$1.406\,\AA\, and $a_{BN}=$1.43\,\AA\, in h-BN which results
in a lattice mismatch ($\delta=\frac{a_{BN}}{a_{CC}}-1$) of 1.7$\%$.
In Figs.~\ref{figMD}(a,b) we depict height deformations in GE due to
the interaction with the h-BN sheet for $\theta=0^{o}$ along (a)
zigzag direction  where $y=0$ and (b) armchair direction where
$x=0$. The corresponding symbols are the analytical results given by
Eq.~(\ref{h2})(see Fig.~\ref{fig1}(a1)). It is seen that the MD
relaxed deformation is  in  good agreement with our analytical
theory. However, the scale parameter $h_0$ was found to be 1.1\,pm
for $\theta=0^o$ and 0.1\,pm for $\theta=0.5^o$ which is one order
of magnitude smaller than found experimentally ~\cite{NAtMat2013}.
As was also noticed recently for the graphene/Cu(111)
system~\cite{graphenecu} this discrepancy can be traced back to the
use of a simple pairwise potential that underestimates $h_0$.
Possibly the use of the yet unavailable three body potential for C-B
and C-N will give a more correct height scale $h_0$. Notice that the
advantage of using MD simulation for studying effects due to MP is
that it allows us to use large nit cells having nano meter size
which is infeasible by ab-initio calculations. The disadvantage of
using our two body potential is that it ignores any direction
dependence.

In summary, we presented a general theory for the strain modulation
in a graphene sheet due to the vdW interaction with a substrate
having a small lattice mismatch $\delta$. The strain results in an
induced pseudo magnetic  field that depends strongly on the
misorientation angle between GE and the substrate. The stress is
distributed non-uniformly following  a Moir\'e pattern. Our
analytical theory was validated qualitatively by using atomistic
simulations that we applied to the GE/BN system. This study realizes
in a natural way the proposal for triaxial stress creation in GE
proposed by F. Guinea \emph{et al.} \cite{NatPhys} using a h-BN
sheet. The obtained Moir\'e pattern agrees with those found
experimentally~\cite{NAtMat2013,geimarXiv,Mashoff} and the induced
gap agrees with recent experiments~\cite{science2013,geimarXiv}.

\textbf{Acknowledgment}: This work was supported by the Flemish
Science Foundation (FWO-Vl) and the Methusalem Foundation of the
Flemish Government.
  M.N.-A was supported by the EU-Marie Curie IIF postdoc Fellowship 299855.

\begin{thebibliography}{15}
\bibitem{Proof}In fact, we performed several large scale atomistic simulations with dimensions larger than 200\AA~and found even for very weak
  vdW interaction (10\,meV) that  the minimum energy
configuration contains out-of-plane deformation with MP structure.
\end{thebibliography}
\end{document}